\documentclass[conference]{IEEEtran}
\usepackage{amsmath,amssymb,amsfonts}
\usepackage{array}
\usepackage[caption=false,font=normalsize,labelfont=sf,textfont=sf]{subfig}
\usepackage{textcomp}
\usepackage{stfloats}
\usepackage{url}
\usepackage{verbatim}
\usepackage{graphicx}
\usepackage{cite}
\usepackage{orcidlink}
\usepackage{colortbl}
\usepackage{xcolor}
\usepackage{makecell} 

\def
\BibTeX{{\rm B\kern-.05em{\sc i\kern-.025em b}\kern-.08em
    T\kern-.1667em\lower.7ex\hbox{E}\kern-.125emX}}
                                                                                                                                                              
\usepackage[ruled, linesnumbered, commentsnumbered, longend]{algorithm2e}
\SetAlCapNameFnt{\footnotesize}
\SetAlCapFnt{\footnotesize}

\usepackage{silence}
\begin{document}

\title{Utilizing Blockchain and Smart Contracts for Enhanced Fraud Prevention and Minimization in Health Insurance through Multi-Signature Claim Processing}

\author{
\IEEEauthorblockN{Md Al Amin, Rushabh Shah, Hemanth Tummala, and Indrajit Ray}
\IEEEauthorblockA{\textit{Computer Science Department, Colorado State University, Fort Collins, Colorado, USA} \\ 
\{Alamin, Rushabh.Shah2, Hemanth.Tummala, Indrajit.Ray\}@colostate.edu}
}

\maketitle

\begin{abstract}
Healthcare insurance provides financial support to access medical services for patients while ensuring timely and guaranteed payment for providers. Insurance fraud poses a significant challenge to insurance companies and policyholders, leading to increased costs and compromised healthcare treatment and service delivery. Most frauds, like phantom billing, upcoding, and unbundling, happen due to the lack of required entity participation. Also, claim activities are not transparent and accountable. Fraud can be prevented and minimized by involving every entity and making actions transparent and accountable. This paper proposes a blockchain-powered smart contract-based insurance claim processing mechanism to prevent and minimize fraud in response to this prevailing issue. All entities—patients, providers, and insurance companies—actively participate in the claim submission, approval, and acknowledgment process through a multi-signature technique. Also, every activity is captured and recorded in the blockchain using smart contracts to make every action transparent and accountable so that no entity can deny its actions and responsibilities. Blockchains' immutable storage property and strong integrity guarantee that recorded activities are not modified. As healthcare systems and insurance companies continue to deal with fraud challenges, this proposed approach holds the potential to significantly reduce fraudulent activities, ultimately benefiting both insurers and policyholders. The average gas costs for smart contract deployment, claim submission, and multi-signature for the Ethereum network: \$80.22, \$20.60, and \$6.47, and for the Optimism network: \$0.35, \$0.089, and \$0.028. They are feasible for the proposed approach.
\end{abstract}


\section{Introduction} \label{sec:introduction} 
Health insurance is important to provide individuals with access to necessary medical care and services without incurring high costs. It supports patients by facilitating treatments, medications, and other essential services. Insurance helps healthcare providers to get continued revenue and business sustainability \cite{Green2001Stakeholder}. This interdependent relationship ensures an operational healthcare system, promoting better health services and financial growth. Insurance fraud introduces a financial burden on the system through false claim payments, which increase premiums. Major healthcare insurance frauds are phantom billing, upcoding, unbundling, kickbacks, identity theft, policyholder fraud, pharmacy fraud, and others \cite{ismail2021healthcare}. These activities vary from billing for non-existent services (\textit{phantom billing}) and manipulating billing codes for higher costs (\textit{upcoding}) to accepting illegal payments for patient referrals (\textit{kickbacks}) and using another identity to receive medical services (\textit{identity theft}). These fraudulent practices contribute to the substantial financial losses incurred by the healthcare system annually, emphasizing the pressing need for a concentrated solution to prevent and minimize fraud \cite{agarwal2023intelligent}.

\begin {figure}[htb]
        \centering
        \includegraphics[width=\linewidth]{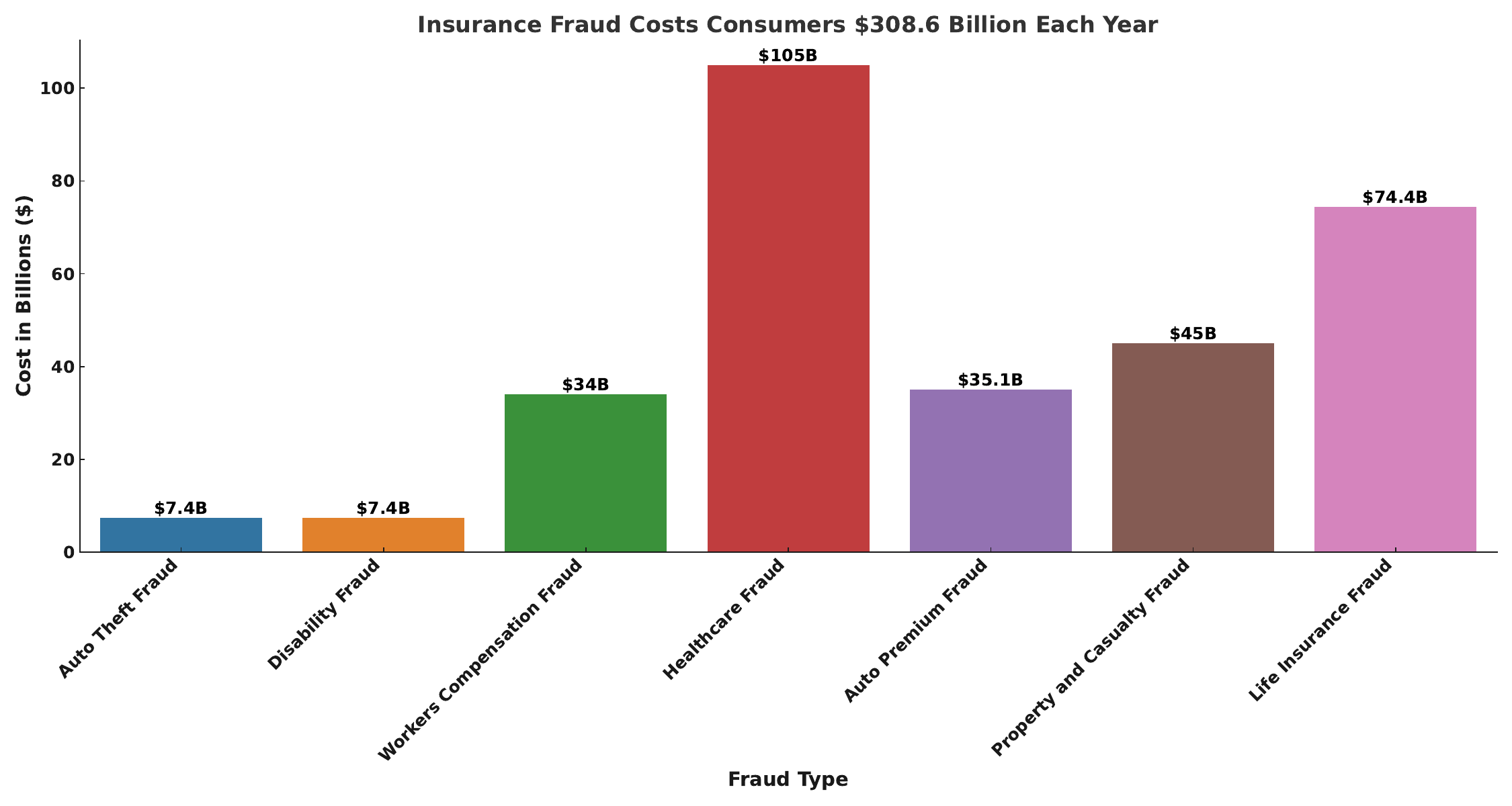}
        \vspace{-2em}
        \caption{Insurance Fraud Statistics \cite{Kilroy2024}} \label{fig:fraud-statistics}
 \end{figure}

The "\textit{Insurance Fraud Statistics 2024}" report by \textit{Forbes Advisor} showcases the extent and financial impact of insurance fraud in the U.S., specifically for healthcare industries \cite{Kilroy2024}. Annually, the U.S. experiences approximated losses of  \$ 306 billion in insurance fraud, translating to roughly \$900 per consumer, primarily due to increased premiums resulting from fraudulent activities. Among various types of insurance fraud, healthcare insurance fraud is the most costly, incurring an estimated \$105 billion in annual losses, as shown in Fig. \ref{fig:fraud-statistics}. Consequently, this results in increased insurance premiums for both individual and corporate policyholders. It weakens public healthcare support, limiting resources availability for legitimate care and services in the healthcare sector. Trust in the healthcare ecosystem is undermined by insurance fraud, which requires a comprehensive and transparent claim processing approach to ensure stakeholders' participation and accountability \cite{agarwal2023intelligent}.

Preventing and minimizing healthcare insurance fraud presents various challenges. Overcoming these requires the combined effort of multiple stakeholders, including patients, providers, and insurance companies. Enforcing patient-level policies and consent is one of the major issues that is frequently disregarded but is vital in this matter. It can lead to unauthorized access to protected health information and fraudulent claims. Moreover, the dependence on centralized hospital systems poses a risk of becoming a single point of failure, emphasizing the need for decentralized audit trails and verifiable claim records to ensure transparent and legitimate claim processing. To prevent and minimize fraud, involving every entity and keeping every step transparent and accountable for timely and proper insurance claim processing is important.

To address the urgency of this problem, this paper proposes a blockchain and smart contract-based three phases—submission, approval, and acknowledgment—healthcare insurance claim processing mechanism. Every involved stakeholder, like patients, providers, and insurance companies, participates in the claim processing to ensure only legitimate claims are processed. Integrating multi-signature transactions at the core of the proposed approach is essential for establishing a decentralized and immutable record of insurance claim interactions. This method ensures that fraudulent activities can be significantly prevented and minimized, as transactions require the consensus of all involved parties—patients, healthcare providers, and insurance companies—before they can be authenticated and finalized on the blockchain. Blockchain is a distributed ledger technology that records transactions across multiple nodes so that the registered transactions cannot be altered. This feature ensures the integrity and transparency of the data once it has been committed to the blockchain \cite{le2021systematic}. 

The proposed approach discourages malicious activities and protects stakeholders' interests by mandating multiple approvals. This simplifies the claims process and increases trust among involved entities. To our knowledge, this work is the first to apply blockchain and smart contract-based multi-signature approach for health insurance claim processing to prevent and minimize fraud. Our contributions include (i) investigating and modeling health insurance fraud mechanisms; (ii) proposing insurance claim processing mechanisms using smart contract-based multi-signature to prevent and minimize fraud by involving every entity in every step; (iii) capturing and storing claim activities on the blockchain network to ensure involved entities are accountable for their actions and activities are immutable and transparent; and (iv) last but not least, developing and deploying smart contracts to execute multi-signature claim transactions and record claim-related activities on the blockchain network.

The remainder of the paper is organized as follows:  Section \ref{sec:related-works} discusses some works related to this paper. Section \ref{sec:insurance-claim-fraud-model} explains the insurance claim and fraud mechanisms with the necessary components. The proposed approach is presented in Section \ref{sec:proposed-appraoch}. The experimental evaluations are given in Section \ref{sec:experiemntal-evaluation}. Section \ref{sec:conclusion} concludes with future research directions.


\section{Related Works} \label{sec:related-works} 
Kapadiya et al. \cite{kapadiya2022blockchain} presented a comprehensive examination of the application of Blockchain and Artificial Intelligence (AI) technologies to detect fraud within healthcare insurance systems. The authors focused on leveraging the immutable and transparent nature of blockchain technology alongside the predictive capabilities of AI to enhance the integrity and efficiency of fraud detection mechanisms. The architecture comprises several layers, including user interaction, data generation, analytical processing, and blockchain integration, each of which plays a crucial role in fraud detection.

Ismail et al. \cite{ismail2021healthcare} proposed a blockchain-based framework called \textit{Block-HI}  to detect multiple fraud scenarios in healthcare industries. \textit{Block-HI }operates on a peer-to-peer network, focusing on one stakeholder—insurance companies—while leaving other stakeholders, such as providers and patients. Their approach is aligned with ours, but in our approach, we focus on all stakeholders rather than a single entity responsible for submitting the claim to the decentralized network. While \textit{Block-HI} presents a comprehensive solution by concentrating on a significant stakeholder, it could overlook the nuanced and individualized tactics other stakeholders employ within the fraud ecosystem.

Al-Quayed et al. \cite{al2023towards} proposed a \textit{Smart Healthcare Insurance Framework for Fraud Detection and Prevention (SHINFDP)} technique, leveraging blockchain, 5G, cloud computing, and machine learning to enhance the security and efficiency of health insurance systems. They highlighted health insurance fraud's financial and operational challenges, detailing various fraudulent activities by healthcare service providers, policyholders, and insurance providers. The empirical study demonstrated the \textit{SHINFDP's} capability to detect healthcare fraud early, showcasing a significant advancement in mitigating fraud risks in health insurance claims processing. However, the machine learning model relies on data that must adapt to changes in \textit{HIPAA} policies, potentially resulting in models becoming outdated for detection purposes or being trained on irrelevant data.

Mendoza-Tello et al. \cite{mendoza2021blockchain} investigated the application of blockchain technology for minimizing healthcare insurance fraud, emphasizing the industry's susceptibility to significant financial losses. They proposed a novel blockchain-based model to improve insurance claim processes' efficiency, integrity, and transparency. The model, structured across four layers, including a peer-to-peer mining pool and a user layer for stakeholders, promises to automate and secure insurance operations, reducing fraud opportunities. Despite its potential, the initial accuracy of medical information and the integration of historical data may limit the model's effectiveness. Further research should focus on collaboration among insurance entities to address these challenges.

Amponsah et al. \cite{amponsah2022novel} presented a novel system integrating machine learning and blockchain technology to tackle healthcare fraud in claim processing. They employed a decision tree classification algorithm, achieving a notable classification accuracy of \textit{97.96\%} and sensitivity of \textit{98.09\%}. This knowledge was then encoded into an \textit{Ethereum} blockchain as smart contracts to enhance insurance data security and transparency in the claim processing system. Despite its successes, the study relied heavily on the accuracy of historical data for training the machine learning model, potentially reducing its ability to detect new or varied fraudulent activities.

Zhang et al. \cite{zhang2022identifying} introduced an innovative framework that integrated blockchain technology and deep learning to address medical insurance fraud. Leveraging a \textit{BERT-LE} model, the authors efficiently automated the detection of fabricated healthcare information. The proposed approach significantly reduces auditors' workload. Furthermore, they proposed a secure and immutable consortium blockchain-based system for storing and managing medical records, ensuring their security, traceability, and auditability. Alhasan et al. \cite{alhasan2021blockchain} developed a blockchain-based system to prevent counterfeit health insurance, enhancing security, transparency, and efficiency by storing data in a decentralized ledger. However, it should be noted that the proposed approach mainly relies on the consensus mechanism of blockchain to detect fraud, which is vulnerable to mitigating the risks of majority attacks.

Kotrakar et al. \cite{kotarkar2022leveraging} present a blockchain-based framework to improve health insurance claim processing. The framework utilizes decentralized storage to ensure the security and immutability of health records. Smart contracts automate the claim approval process, minimizing human intervention and accelerating transaction rates. The system architecture comprises insurance companies, hospitals, and policyholders interacting through a secure blockchain network. Endorsers/validators verify transactions to maintain data integrity. However, a major issue is that the system primarily focuses on logging information, leaving room for potential fraud.

The papers mentioned above summarize how they have used machine learning-based techniques to detect fraud where data is stored in the blockchain. However, to prevent and minimize fraud, it is important to involve every entity and keep every step transparent and accountable in the timely and proper processing of insurance claims. This paper proposes a blockchain and smart contract-based approach for preventing and minimizing fraud claims. Where every involved entity approves corresponding claim activities—submission, approval, and acknowledgment—through multi-signature transactions. Activities information is captured and recorded on the blockchain to provide immutable and transparent claim-related data.


\section{Insurance Claim and Fraud Mechanism}\label{sec:insurance-claim-fraud-model}
\subsection{Ideal Insurance Claim Workflow}
\begin{figure*}[tb]
        \centering
        \includegraphics[width=0.92\linewidth]{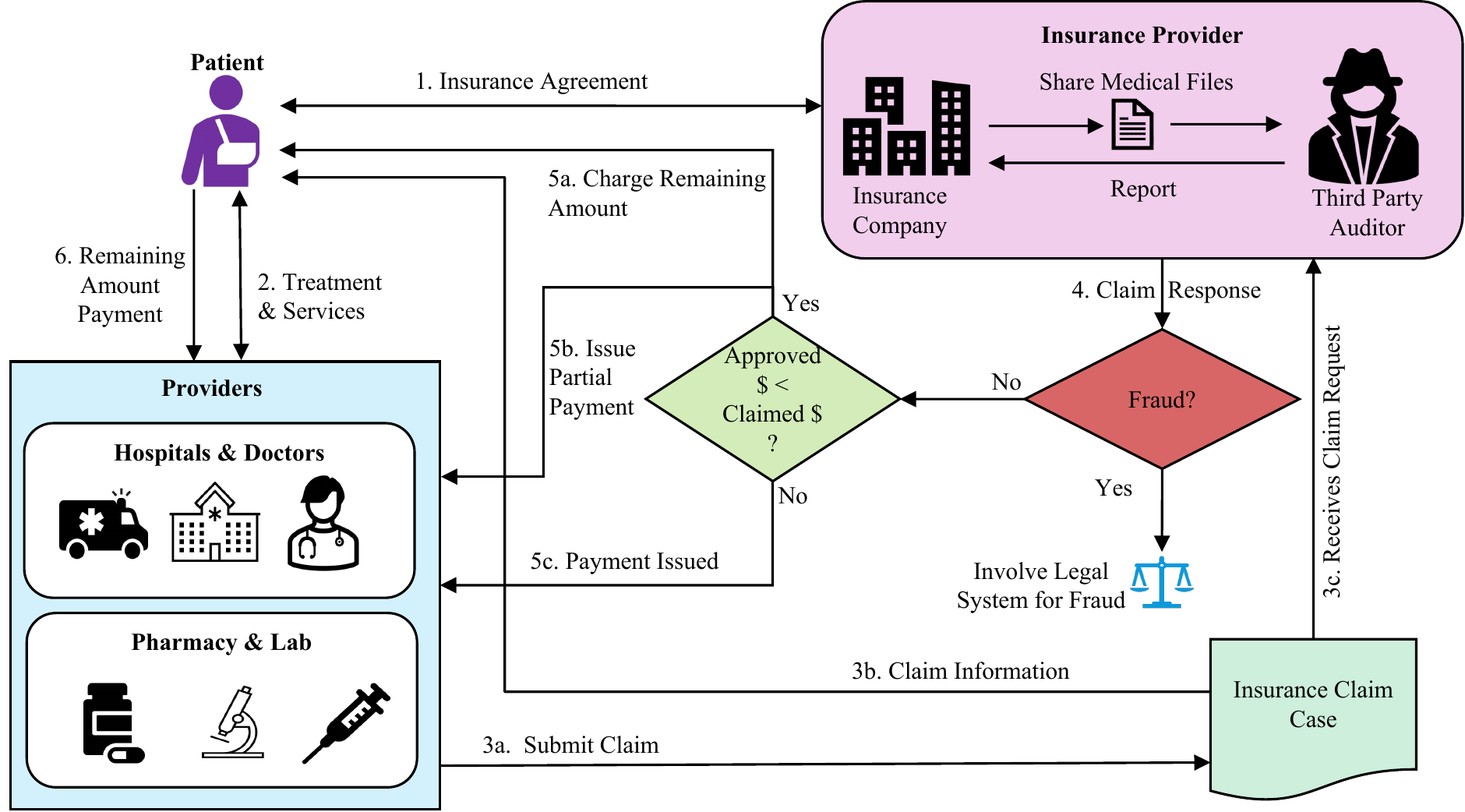}
        \vspace{-2em}
        \caption{Insurance Claim Processing Model} \label{fig:insurnace-claim-and-fraud-mechanism}
    \end{figure*}

The mechanism of dealing with insurance claims is presented in Fig. \ref{fig:insurnace-claim-and-fraud-mechanism}. \textit{Step 1}: Patient enters the insurance contract. \textit{Step 2}: healthcare provider treats patient. \textit{Step 3a,3b}: Provider sends claim information to the insurer and the patient. \textit{Step 3c}: The insurance company processes the claim information. \textit{Step 4}: The insurance provider responds to the claim. If the claim is approved: \textit{Step 5a}: In case the amount claimed is greater than the authorized amount, then in such cases the amount greater than the authorized amount is going to be charged to the patient by the insurance provider. \textit{Step 5b}: Provider gets partial payment. \textit{Step 5c}: Full payment if the claim is fully approved. \textit{Step 6}: Then the patient pays the rest of the remaining amount to the provider. If fraud is suspected during the claim review (\textit{Step 4}), the case is referred to include the legal system. The diagram portrays the interrelations between actors in insurance claims processing and the significance of meticulousness and compliance with the procedures.

\subsection{Health Insurance Fraud Classification}
Fraud happens in many known and unknown ways. The following section discusses the most common and widely recognized health insurance fraud and its mechanisms and consequences.

\subsubsection{Phantom Billing}
In this fraud, healthcare providers claim compensation for non-existent services or procedures \cite{rath2023healthcare}. Examples include billing for medical consultations or diagnostic tests that never took place. When a patient visits, a healthcare provider receives standard examinations and tests. Post-visit, healthcare providers create false records, documenting services never provided. The submission of fraudulent claims, which involve a combination of genuine and fabricated services, exploits the system's complexity and the company's trust in providers. Unaware of the fraud, the insurance company processes and pays these claims, leading to profit for the healthcare provider. 

\subsubsection{Upcoding}
This billing code manipulation involves healthcare providers billing for more expensive treatments or services than were provided. For instance, this may involve billing for a complex medical procedure when a less intricate one was actually performed \cite{milcent2021downcoding}. As a result, insurance companies reimburse based on inflated claims, leading to unnecessary financial losses. Due to inflated billing, patients may also face higher co-payments, deductibles, or depletion of their annual insurance benefits.

\subsubsection{Unbundling}
Healthcare providers sometimes separate a comprehensive service, typically billed as a single package, into its components for billing purposes \cite{li2020financial}. This disaggregation often results in a cumulative charge more significant than the bundled rate, unjustly increasing the insurance payout. Providers achieve this by listing each component of a bundled procedure as separate billable items, thereby inflating the total charge. Consequently, insurance companies face higher reimbursement rates, leading to increased healthcare costs. Due to the disaggregation of services, this practice results in higher out-of-pocket expenses for patients.

\subsubsection{Kickbacks}
It involves healthcare professionals receiving unauthorized benefits in exchange for patient referrals, prescription of specific drugs, or endorsements of particular treatments \cite{chen2020recommendations}.  This compromises medical ethics and the integrity of medical decisions by prioritizing financial incentives over patient care, potentially leading to unnecessary treatments or medication prescriptions. Consequently, patients may receive unnecessary or more expensive treatments due to these kickback-driven referrals, undermining their trust and affecting their health and financial well-being.

\subsubsection{Identity Theft}
This fraud involves individuals illicitly acquiring another person's personal information to access medical services or medications, often at the expense of the victim's insurance coverage, leading to false medical records and financial loss \cite{goel2020medical}. For the victims, this unauthorized use of their insurance can result in financial liabilities and potential loss of coverage. Resolving medical identity theft's consequences can also be complex and time-consuming.

\subsubsection{Policyholder Fraud}
Individuals commit this fraud by providing inaccurate or incomplete information on insurance claims \cite{saldamli2020health}. This deceit often involves concealing pre-existing conditions or falsifying health details to gain or increase coverage, leading to undue benefits. By providing false information on applications, individuals affect premium calculations. This misrepresentation skews risk assessments and premium rates, potentially increasing costs for all policyholders.

\subsubsection{Pharmacy Fraud}
Pharmacies may engage in deceptive practices such as billing for higher-priced medications than dispensed, charging for prescriptions that were not provided, or making fraudulent reimbursement claims, undermining healthcare systems' trust and financial stability \cite{baranek2018tricare}. These discrepancies include overcharging for medications, billing for undispensed drugs, and submitting false claims for reimbursement. As a result, insurance companies face financial losses by reimbursing these fraudulent claims.

\begin{table}[tp]
    \centering
    \scriptsize
    \caption{Impact of Different Types of Fraud on Various Stakeholders} \label{table:fraud_impact}
    \vspace{-1em}
    \rowcolors{2}{teal!25}{gray!20}
    \begin{tabular}{|c|c|c|c|} \hline
      \rowcolor{teal!60} \textbf{Fraud Type} & \textbf{Provider} & \textbf{Insurer} & \textbf{Patient}\\ 
        \hline
        Phantom Billing & Malicious & Non-Malicious & Non-Malicious\\ 
        \hline
        Upcoding & Malicious & Non-Malicious & Non-Malicious \\ 
        \hline
        Unbundling & Malicious & Non-Malicious & Non-Malicious\\ 
        \hline
        Kickbacks & Malicious & Non-Malicious & Non-Malicious\\ 
        \hline
        Identity Theft & Non-Malicious & Non-Malicious & Malicious\\ 
        \hline
        Policyholder Fraud & Non-Malicious & Non-Malicious & Malicious\\ 
        \hline
        Pharmacy Fraud & Malicious & Non-Malicious & Non-Malicious\\ 
        \hline
    \end{tabular}
\end{table}

\subsection{Fraud Insurance Claim Process} 
In the healthcare sector, the prevalence of fraud poses significant challenges to all stakeholders, including healthcare providers, insurance companies, and patients. However, not every entity is involved in all steps of fraudulent activities, and accountability is often lacking, leading to a lack of transparency in actions. The analysis from Table \ref{table:fraud_impact} reveals that few stakeholders are malicious, yet they account for most fraud cases. This highlights a critical weakness: malicious actors can exploit the system without full accountability and transparency, affecting non-malicious entities. Furthermore, detecting fraud becomes nearly impossible if the majority are malicious in a group of stakeholders and only one is honest.

By leveraging blockchain technology and multi-signature transactions, the healthcare sector can significantly reduce the risk of fraud. This approach enhances the integrity of financial transactions and fosters a climate of trust among all stakeholders, which is crucial for the effective delivery of healthcare services. Multi-signature in this context means that multiple entities have used their private keys to sign the transactions, making them responsible for doing so.

\section{Proposed Approach}\label{sec:proposed-appraoch}
\subsection{Approach Overview}
\begin{figure*}[tb]
        \centering
        \includegraphics[width=.85\linewidth]{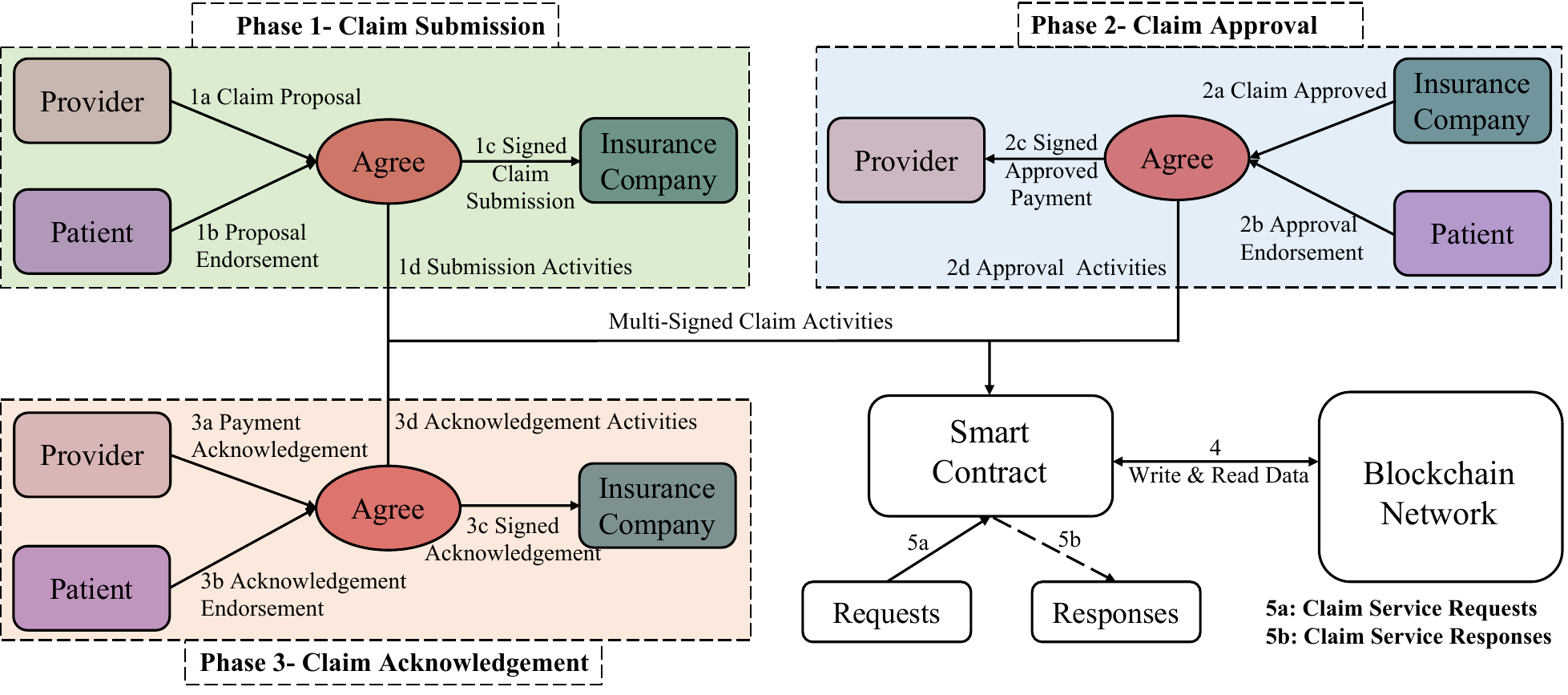}
        \vspace{-1em}
       \caption{Proposed Multi-Signature Based Insurance Claim Processing Mechanism} \label{fig:proposed-claim-processing-approach}
\end{figure*}

\begin{figure*}[tb]
        \centering
        \includegraphics[width=\linewidth]{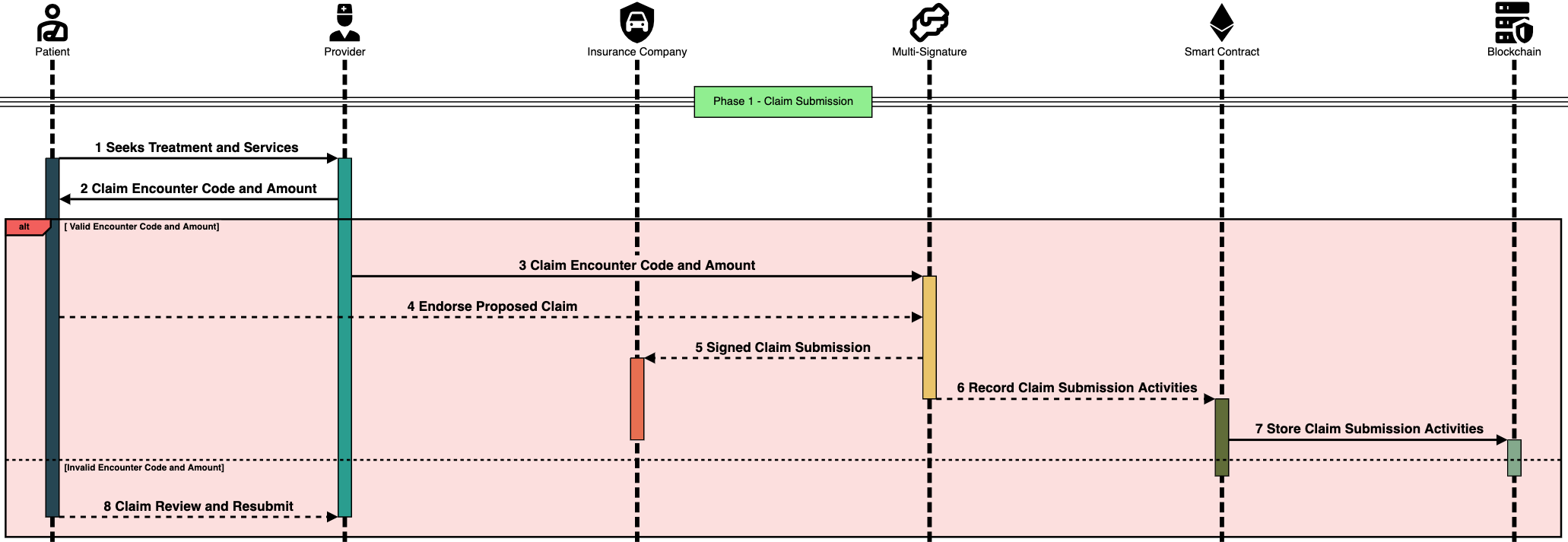}
        \vspace{-2em}
       \caption{Proposed Insurance Claim Submission Phase Sequence Diagram} \label{fig:proposed-claim-submission-sequence}
\end{figure*}

In the proposed claim processing approach, there are three phases: \textit{(i) Phase 1- Claim Submission; (ii) Phase 2- Claim Approval;} and \textit{(iii) Phase 3- Claim Acknowledgement}. Fig. \ref{fig:proposed-claim-processing-approach} shows all three phases. In \textit{Phase 1}, the provider submits an insurance claim for providing treatment and services to the patient. To make the process transparent and accountable, the corresponding patient and the provider must approve the submitted claim. Fig. \ref{fig:proposed-claim-submission-sequence} shows the sequence diagram for this phase. In \textit{Phase 2}, the insurance company processes the claim submitted by the provider according to the policy contract with the patient. After processing the claim, the insurance company approved an amount that may be equal to or less than the submitted amount. At this stage, the insurance company and the patient must endorse the approved amount. Then, the amount is credited to the provider's bank account.

After receiving the amount against a submitted claim, the provider and patient agree and submit the acknowledgment to the insurance company in \textit{Phase 3}.  Providers and insurance companies can't process false claim since the patient is involved in all the processes. All activities are recorded and stored in the blockchain. Smart contacts are deployed to capture the transactions signed by the corresponding entities using their private keys through web or mobile application services. This makes denying any user the submitted transaction impossible unless their credentials are compromised. Each step integrates multi-signature verification from the involved stakeholders to create a fraud-tolerant and reliable healthcare claim process using blockchain technology. This ensures that every stage of the claim is transparent, auditable, and secure.

\begin{table}[tb]
    \centering
        \caption{Definition of Notations for Algorithms (\ref{alg:phase1} - \ref{alg:phase3})} \label{tab:definition_of_notations}
        \vspace{-1em}
        \scriptsize
         \rowcolors{2}{purple!25}{gray!20}
    \begin{tabular}{|l|l|} \hline
        \rowcolor{purple!50}   \textbf{Notation} & \textbf{Description} \\
            \hline 
                $\mathbb{ID}_{Insurnace Policy}$  & Patient and insurance company agreement \\
            \hline 
                $\mathbb{ID}_{Patient}$           & Who seeks treatment \& services from providers \\
            \hline 
                $\mathbb{ID}_{Provider}$          & Who provides treatment \& services, expect payment \\
            \hline 
                $\mathbb{ID}_{Insurance}$         & Who processes claims submitted by providers \\
            \hline 
                $\mathbb{BN}$         & Blockchain network \\
            \hline
                $\mathbb{SC}$         & Smart contract\\
            \hline
    \end{tabular}
\end{table}

In addition to the above three phases, there is an additional phase called \textit{Pre-Phase: Insurance Premium Agreement Initialization}. Although, the proposed approach doesn't provide any mechanism for this phase. Instead, it discusses essential activities to ensure patients have proper insurance coverage for treatment and services. In this stage, patients, also known as policyholders, submit the required information, such as medical history, current conditions, and other relevant health information, to the insurance company for processing. The submitted data undergoes a verification process by healthcare providers or institutions designated by the insurance company to ensure accuracy and authenticity.  After the data review, an insurance contract is established between the insurance company and the patient on benefits, scopes, and premiums. This contract also encodes the terms of data use, privacy protections, and the responsibilities of both parties. Policyholders set up emergency contacts and manage consent for how their medical data is used in interactions with healthcare providers. This includes providing permissions for treatment procedures, prescriptions, and laboratory tests and ensuring full control over their personal medical information.

\subsection{Phase 1- Claim Submission}
This phase is initiated when a patient receives treatment and services from providers, including primary care physicians, hospitals, pharmacies, dental clinics, laboratories, etc. Healthcare providers create medical cases that summarize the services and treatments provided. Each case includes health codes that categorize the medical services rendered. The total charge for the given treatment and services is calculated. If a copayment occurs, that amount is deducted from the total bill and charged directly to the patient's account. The remaining balance is sent to the patient for review and submission with the encounter code and corresponding amount to the insurance company. \textit{Left-Upper} box in Fig. \ref{fig:proposed-claim-processing-approach} shows the claim submission process. The provider submits an insurance claim in \textit{Step 1a}. The patient reviews and endorses the submitted claim in \textit{Step 1b}. \textit{Steps 1a} and \textit{1b} together make a multi-signature transaction that is agreed upon by the provider and the patient. If there are issues, like the wrong amount for a given service, the patient and provider work together to solve them. 

The patient knows the compensation for any encounter code or service from the \textit{Pre-Phase}. This helps the patient screen out any unexpected claim submission. When the provider and patient sign a proposed claim, then the proposed claim is submitted in \textit{Step 1c} to the insurance company for further processing. The activities of this phase are captured and recorded in the blockchain network through a smart contract in \textit{Step 1d}. Provider and patient use their corresponding private key to sign the transaction or claim using a wallet like \textit{Metamask} or \textit{Coinbase} \cite{han2021efficient}, ensuring that all services are accurately represented and authorized. This signing ensures that they can not deny the claim submission activities later, which ensures accountability. Algorithm \ref{alg:phase1} contains the step-by-step instructions for this phase. It is assumed that the patient gives the required consent for the provider to share their protected health information with the insurance company.

\RestyleAlgo{ruled}
\SetKwComment{Comment}{/* }{ */}
\begin{algorithm}[tb]
\scriptsize
    \SetKwInOut{KwData}{Input}
    \SetKwInOut{KwResult}{Output}
    \SetKw{KwBy}{by}
\DontPrintSemicolon

\caption{Phase 1- Claim Submission}\label{alg:phase1}

\KwData{ (i) initial claim with encounter code ($EC$) and amount ($AMT$)}
\KwResult{ (i) submitted claim $\mathbb{Tn}_{Submitted}$ with encounter code and amount}
\textcolor{blue}{\Comment*[r]{Providers submit insurance claim with \textit{Encounter Code} and \textit{Amount} for patients}}
\;

 \textbf{Claim Review} \vfill
     \quad  \textit{(a)} $\mathbb{C}_{EC}$  \textcolor{blue}{\Comment*[r]{claim with encounter code list }}
     \quad  \textit{(b)} $\mathbb{C}_{AMT}$  \textcolor{blue}{\Comment*[r]{amount for included encounter code}}
     \quad  \textit{(c)} $\mathbb{Tn}_{Submitted} = []$  \textcolor{blue}{\Comment*[r]{valid $EC$ and $AMT$}}\;

                        \For{$i \gets {\mathbb{C}_{EC}}_{Start}$ \KwTo ${\mathbb{C}_{EC}}_{End}$ \KwBy $1$}{
                                \eIf{$\zeta ({\mathbb{C}_{EC}}_{i},{\mathbb{C}_{AMT}}_{i})$ is in $\mathbb{ID}_{Insurnace Policy}$}{
                                        \textcolor{blue}{\Comment*[r]{amounts are within the coverage range}}
                                        \textcolor{blue}{\Comment*[r]{initial review by patient and provider}}
                                       \quad  \textit{(i)} $\mathbb{Tn}_{Submitted} \gets \mathbb{Tn}_{Submittedd} + ({\mathbb{C}_{EC}}_{i},{\mathbb{C}_{AMT}}_{i})$\;
                                }{
                                      \textbf{\textit{ Error: }}Invalid ${\mathbb{C}_{EC}}_{i}$ and ${\mathbb{C}_{AMT}}_{i}$\;
                                }
                        }
 \;
 \textbf{Claim Submission} \vfill
    \eIf{Sign($\mathbb{Tn}_{Submitted}$, $\mathbb{ID}_{Provider}$) \&\& Sign($\mathbb{Tn}_{Submitted}$, $\mathbb{ID}_{Patient}$) == True}{
                                 \textcolor{blue}{\Comment*[r]{Patient knows all claim acitivities}}
                     \quad \textit{(i)} Submit $\mathbb{Tn}_{Submitted}$ to $\mathbb{ID}_{Insurance}$\;
                     \quad \textit{(ii)} Add activities to the Blockchain Network through Smart Contract
                            \textcolor{blue}{\Comment*[r]{Immutable, transparent, accountable actions}}\;
            }{
                                      \textbf{\textit{ Error: }} $\mathbb{Tn}_{Submitted}$ is not submitted to $\mathbb{ID}_{Insurance}$
            }
\end{algorithm}

\subsection{Phase 2- Claim Approval}
The insurance company receives the healthcare case file, which includes all relevant medical data and documentation related to the services provided to the patient. The claim is scrutinized in-house or by third-party auditors to validate its legitimacy and accuracy against standard medical service codes and policy terms. A report detailing the verification findings and recommendations is compiled. Based on this report, the insurance company either (i) approves the claim and processes payment, or (ii) denies the claim, and/or takes legal action if fraud is suspected. After the initial claim approval, the insurance company shares it with the policyholder. If the decision satisfies the insurance coverage agreed upon in the \textit{Pre-Phase}, the patient consents. In \textit{Steps 2a }and \textit{2b}, the insurance company and the patient endorse the submitted claim through multi-signature. In \textit{Step 2c}, the insurance company will release the payment upon claim approval. The decision on the claim is updated on a blockchain in \textit{Step 2d}, requiring a multi-signature transaction from the patient and the healthcare provider to ensure authenticity and agreement on the documented outcome.  The detailed process of claim approval is listed in Algorithm \ref{alg:phase2}. If the decision is unsatisfactory or violates major premium benefits for the patient or healthcare provider, they can seek legal justification through formal dispute resolution mechanisms.

\RestyleAlgo{ruled}
\SetKwComment{Comment}{/* }{ */}

\begin{algorithm}[tb]
\scriptsize
    \SetKwInOut{KwData}{Input}
    \SetKwInOut{KwResult}{Output}
    \SetKw{KwBy}{by}
\DontPrintSemicolon

\caption{Phase 2- Claim Approval}\label{alg:phase2}

\KwData{ (i) submitted claim $\mathbb{Tn}_{Submitted}$ with encounter code and amount}
\KwResult{ (i) approved claim $\mathbb{Tn}_{Approved}$ with encounter code and amount}\;

 \textbf{Claim Review and Approve} \vfill
        \quad \textit{(a)}  $\mathbb{Tn}_{Submitted}$  \textcolor{blue}{\Comment*[r]{submitted insurance claim }}
       \quad \textit{(b)}  $\mathbb{Tn}_{Approved} = []$  \textcolor{blue}{\Comment*[r]{approved insurance claim}}\;

        \For{$j \gets {\mathbb{Tn}_{Submitted}}_{Start}$ \KwTo ${\mathbb{Tn}_{Submitted}}_{End}$ \KwBy $1$}{
                                \eIf{$\mathbb{Tn}_{{Submitted}_j}$ is in $\mathbb{ID}_{Insurnace Policy}$}{
                                            \textcolor{blue}{\Comment*[r]{amounts are within the coverage range}}
                                        \quad \textit{(i)} $\mathbb{Tn}_{Approved} \gets \mathbb{Tn}_{Approved} + \mathbb{Tn}_{{Submitted}_j}$\;
                                            \textcolor{blue}{\Comment*[r]{approve the amount as submitted}}
                                }{
                                       \quad \textit{(ii)} $ \varphi_{\mathbb{Tn}_{{Submitted}_j}} \gets \mathbb{Tn}_{{Submitted}_j} \pm \delta$ \;
                                                \textcolor{blue}{\Comment*[r]{adjust $\mathbb{Tn}_{{Submitted}_j}$ by $\delta$ amount}}
                                       \quad \textit{(iii)} $\mathbb{Tn}_{Approved} \gets \mathbb{Tn}_{Approved} + \varphi_{\mathbb{Tn}_{{Submitted}_j}} $\;
                                                \textcolor{blue}{\Comment*[r]{approve $\varphi_{\mathbb{Tn}_{{Submitted}_j}}$ }}
                                }
                        }
\;
\textbf{Claim Payment} \vfill
    \eIf{Sign($\mathbb{Tn}_{Approved}$, $\mathbb{ID}_{Insurance}$) \&\& Sign($\mathbb{Tn}_{Approved}$, $\mathbb{ID}_{Patient}$) == True}{
                            \textcolor{blue}{\Comment*[r]{Patient- no objection and agrees on amount}}
                    \quad \textit{(i)} Submit $\mathbb{Tn}_{Approved}$ to $\mathbb{ID}_{Provider}$\;
                    \quad \textit{(ii)} Add activities to the Blockchain Network through Smart Contract
                        \textcolor{blue}{\Comment*[r]{Immutable, transparent, accountable actions}}\;
            }{
                                      \textbf{\textit{ Error: }} $\mathbb{Tn}_{Approved}$ is not submitted to $\mathbb{ID}_{Provider}$
            }
\end{algorithm}

\subsection{Phase 3- Claim Acknowledgement}
Once the insurance company has approved the claim, payment is processed and disbursed to the healthcare provider. This step marks the beginning of the claim acknowledgment phase. The healthcare provider initiates the claim acknowledgment phase upon receiving the payment. This involves the formal recognition that payment has been received. If the approved amount is not the same as the one submitted, the remaining balance is charged to the patient. The provider sends payment information to the patient for review. From \textit{Phase-2}, the patient already knows the amount approved by the insurance company for a submitted claim. The patient agrees to the terms of the approved and received amount, which is equal to a specific claim. Provider and patient together sign the acknowledgment through multi-signature and send it to the insurance company. \textit{Steps 3a, 3b,} and \textit{3c} show this process in Fig. \ref{fig:proposed-claim-processing-approach}. Following the successful multi-signature acknowledgment, all activities related to this phase are captured in \textit{Step 3d} and recorded on the blockchain through the smart contract to reflect that the payment has been acknowledged.  This update is vital for tracking the progress and current status of the claim throughout the healthcare system. Algorithm \ref{alg:phase3} includes step-by-step instructions for this phase.

\RestyleAlgo{ruled}
\SetKwComment{Comment}{/* }{ */}

\begin{algorithm}[tb]
\scriptsize
    \SetKwInOut{KwData}{Input}
    \SetKwInOut{KwResult}{Output}
    \SetKw{KwBy}{by}
\DontPrintSemicolon
\caption{Phase 3- Claim Acknowledgement}\label{alg:phase3}

\KwData{ (i) received claim $\mathbb{Tn}_{Received}$ with encounter code and amount}
\KwResult{ (i) formal acknowledgement ($\mathbb{ACK}$) on $\mathbb{Tn}_{Received}$}

\;
 \textbf{Review Received Payment} \vfill

       \quad \textit{(a)}  $\mathbb{Tn}_{Submitted}$  \textcolor{blue}{\Comment*[r]{submitted insurance claim }}
       \quad \textit{(b)}  $\mathbb{Tn}_{Received}$  \textcolor{blue}{\Comment*[r]{received insurance claim}}
       \quad \textit{(c)}  $\mathbb{Tn}_{Remaining} = []$  \textcolor{blue}{\Comment*[r]{claim amount not covered}}\;

        \For{$k \gets \mathbb{Tn}_{Received_{Start}}$ \KwTo ${\mathbb{Tn}_{Received}}_{End}$ \KwBy $1$}{
                                \eIf{($\mathbb{Tn}_{{Received}_k} < \mathbb{Tn}_{{Submitted}_k}$)}{
                                            \textcolor{blue}{\Comment*[r]{Charge remaining amount to patient }}\;
                                       \quad \textit{(i)} $\mathbb{Tn}_{Remaining} \gets \mathbb{Tn}_{Remaining} + (\mathbb{Tn}_{{Submitted}_k} - \mathbb{Tn}_{{Received}_k})$\;
                                }{
                                    \quad \textit{(ii)} $\mathbb{Tn}_{{Submitted}_k}$ is covered by $\mathbb{Tn}_{{Received}_k}$ \;
                                    \quad \textit{(iii)} $\mathbb{Tn}_{Remaining} \gets \mathbb{Tn}_{Remaining}$ + 0\;
                                }
                        }
\;
\textbf{Claim Acknowledgement} \vfill
    \eIf{Sign($\mathbb{Tn}_{Received}$, $\mathbb{ID}_{Provider}$) \&\& Sign($\mathbb{Tn}_{Received}$, $\mathbb{ID}_{Patient}$) == True}{
                \textcolor{blue}{\Comment*[r]{Provider, Patient agree received payment}}
                    \quad \textit{(i)} Submit $\mathbb{ACK}$ on $\mathbb{Tn}_{Received}$ to $\mathbb{ID}_{Insurance}$\;
                    \quad \textit{(ii)} Send $\mathbb{Tn}_{Remaining}$  to $\mathbb{ID}_{Patient}$\;
                    \quad \textit{(iii)} Add activities to the Blockchain Network through Smart Contract
                            \textcolor{blue}{\Comment*[r]{Immutable, transparent, accountable actions}}\;
            }{
                                      \textbf{\textit{ Error: }} $\mathbb{ACK}$ is not sent to $\mathbb{ID}_{Insurance}$
            }
\end{algorithm}

\section{Experimental Evaluation  } \label{sec:experiemntal-evaluation}
We developed a smart contract in our experimental evaluation and deployed it on two different Ethereum networks: the \textit{Ethereum} mainnet and \textit{Optimism}, a Layer 2 solution. Layer 2 solutions in blockchain technology are secondary protocols that enhance scalability and efficiency by managing transactions off the main chain through mechanisms like state channels, rollups, and sidechains \cite{gudgeon2020sok}. The contract was meticulously crafted during the pre-phase, ensuring the conditions and requirements for creating a reliable smart contract were met. This included the payment of a one-time fee required for the initial deployment of the contract.

\begin{figure*}[htb]
    \centering
    \includegraphics[width=.95\textwidth]{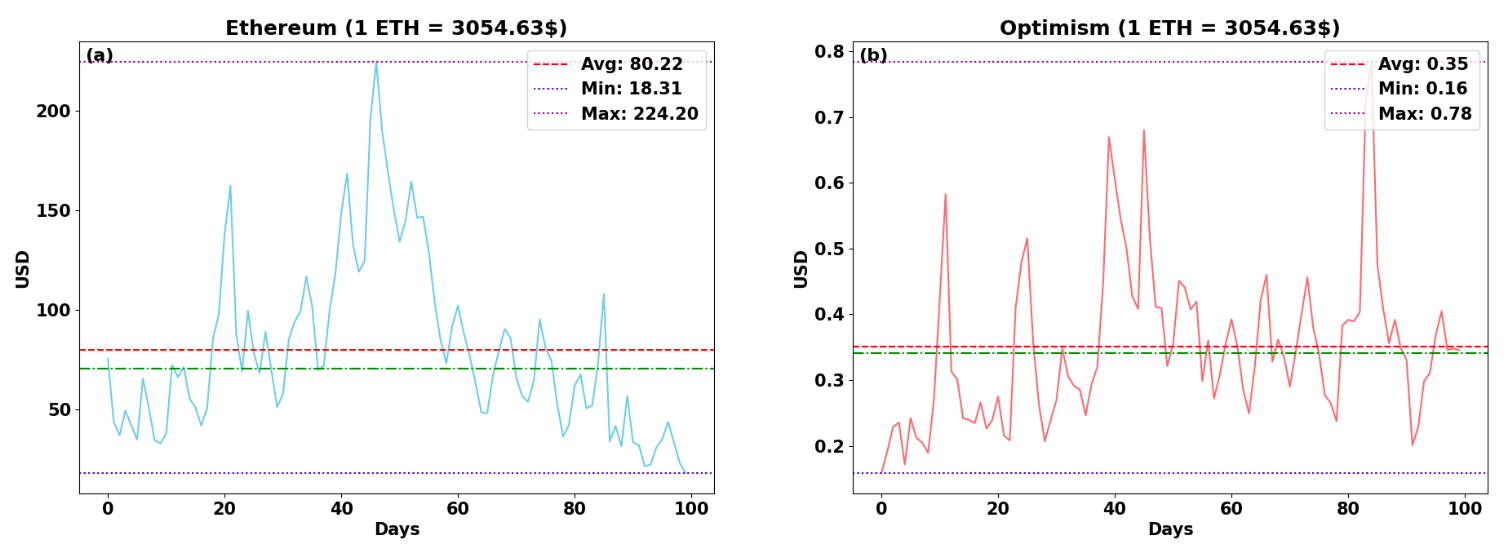}
    \vspace{-1em}
    \caption{Smart Contract Deployment Cost for Ethereum and Optimism Networks} \label{fig:contract-deployment-cost}
\end{figure*}

\begin{figure*}[htb]
    \centering
    \includegraphics[width=.95\textwidth]{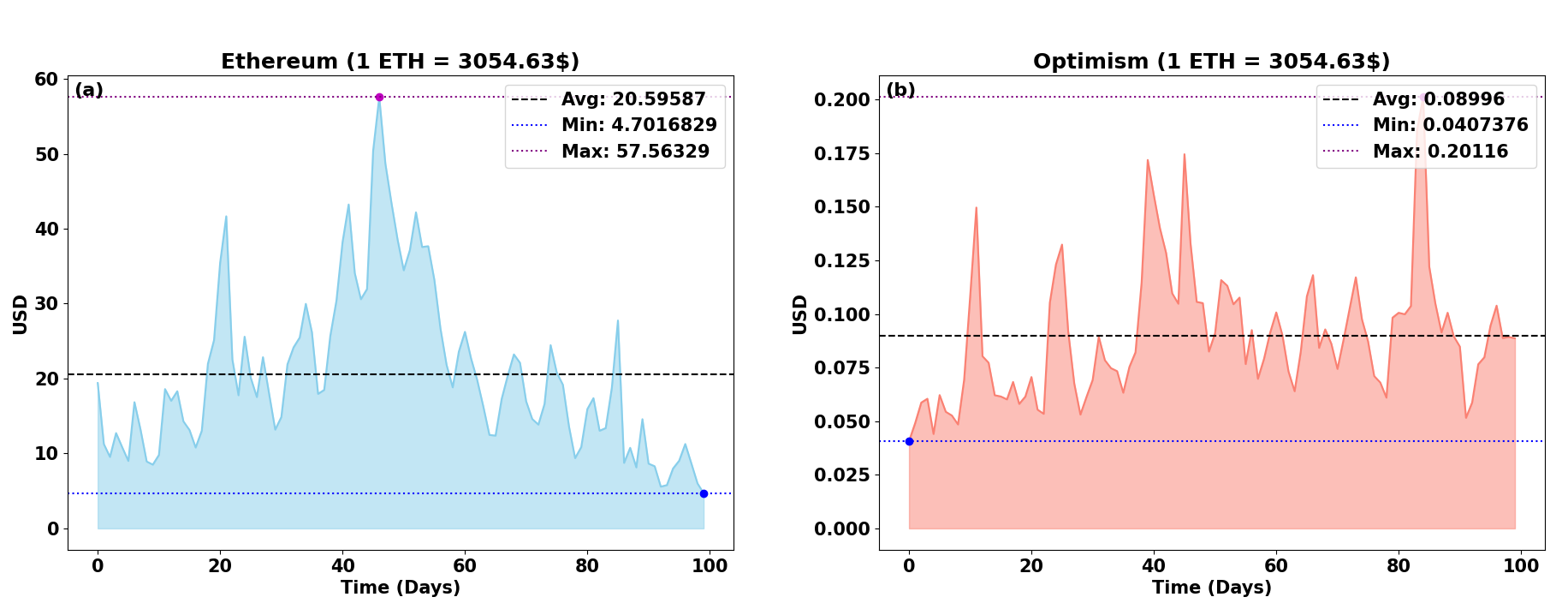}
     \vspace{-1em}
    \caption{Claim Submission Cost for Ethereum and Optimism Networks} \label{fig:claim_sub}
\end{figure*}

\begin{figure*}[htb]
    \centering
    \includegraphics[width=.95\textwidth]{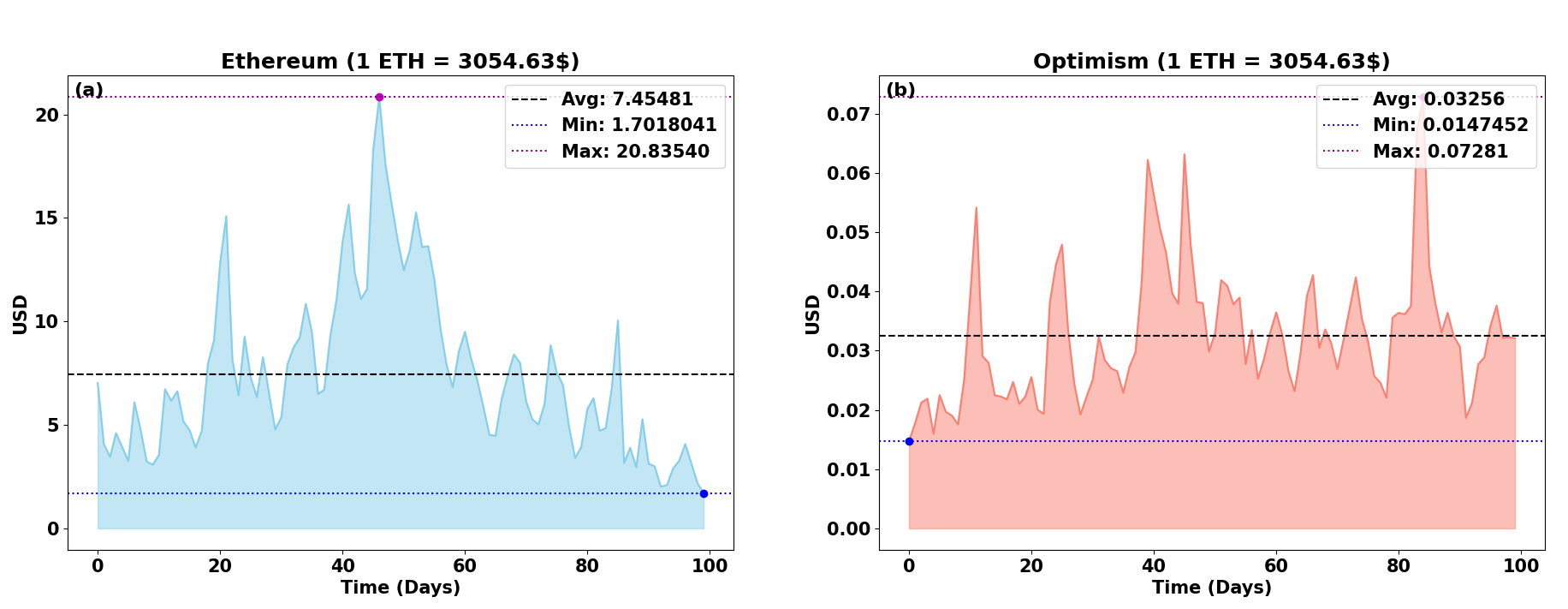}
     \vspace{-1em}
    \caption{Multi-Signature Cost for Ethereum and Optimism Networks} \label{fig:multisign}
\end{figure*}

To assess the financial implications of deploying such a contract, particularly from the perspective of an insurance model, we conducted a comprehensive cost analysis over \textit{100} days. This analysis involved testing the deployment expenses to determine the average cost associated with the contract's deployment. The data collected from this testing phase provided valuable insights into the cost-effectiveness and financial feasibility of deploying smart contracts on the \textit{Ethereum} and the \textit{Optimism} network, aiding in strategic decision-making for future deployments. The price of \textit{Optimism} variation is less dramatic than \textit{Ethereum}, as shown in Fig. \ref{fig:contract-deployment-cost}, indicating more stable and predictable costs.

\subsection{Smart Contract Deployment Cost}
Fig. \ref{fig:contract-deployment-cost} shows the smart contract deployment cost for the \textit{Ethereum} mainnet and \textit{Optimism} network. The prices fluctuate significantly, with a maximum of \textit{\$224.21} and a minimum of \textit{\$18.31} for the \textit{Ethereum} as shown in Fig. \ref{fig:contract-deployment-cost}a. The average deployment cost over the \textit{100} days is about \textit{\$80.22}. The graph shows several spikes, suggesting periods of high gas prices, possibly due to network congestion. Fig. \ref{fig:contract-deployment-cost}b indicates that the costs for the \textit{Optimism} network are generally lower than on the \textit{Ethereum}, with values ranging from \textit{\$0.16} to \textit{\$0.78}. The average cost is much lower at \textit{\$0.35}. 

\subsection{Claim Submission Cost}
After deploying smart contracts, they are called to perform defined operations. For reading operations, it doesn't cost any gas or money. There is a mandatory cost for performing any write operations, depending on the size of the operations. For submitting claims, the associated costs are presented in Fig. \ref{fig:claim_sub} for both networks. Fig. \ref{fig:claim_sub}a shows the prices for the \textit{Ethereum} mainnet with a maximum of \textit{\$57.56} and a minimum of \textit{\$4.70}. The prices fluctuate significantly. The average transaction cost over the \textit{100} days is about\textit{ \$20.60}. The graph shows several spikes, suggesting periods of high gas prices, possibly due to network congestion. The costs for \textit{Optimism} are depicted in Fig. \ref{fig:claim_sub}b, which are generally lower than on the \textit{Ethereum} network, with values ranging from \textit{\$0.040} to \textit{\$0.201}. The average cost is much lower at \textit{\$0.089}. 

\subsection{Multi-Signature Cost}
Every action or transaction for claim submission, claim processing, and payment acknowledgment must be signed by the two entities, as shown in Fig. \ref{fig:proposed-claim-processing-approach}. It costs for each multi-signature operation. Fig. \ref{fig:multisign} shows the \textit{Ethereum} and \textit{Optimism} blockchain network costs. The prices fluctuate significantly, with a maximum of \textit{\$18.08} and a minimum of \textit{\$1.48} for \textit{Ethereum}, as noted in Fig. \ref{fig:multisign}a.  The average transaction cost over the \textit{100} days is about \textit{\$6.47}. The graph shows several spikes, suggesting periods of high gas prices, possibly due to network congestion. Fig. \ref{fig:multisign}b shows the cost for \textit{Optimism}, which is lower than on the \textit{Ethereum}, with values ranging from \textit{\$0.013} to \textit{\$0.063}. The average cost is much lower at \textit{\$0.028}.

\section{Conclusion and Future Directions} \label{sec:conclusion}
This paper explores the integration of blockchain and smart contracts-based multi-signature claim processing as a solution to detect and minimize healthcare insurance fraud. It highlights the critical need to address the significant financial losses the healthcare system incurs annually due to fraudulent activities. The proposed blockchain-based system is designed to provide a transparent, secure, and efficient insurance claim processing mechanism. Multi-signature-oriented claim processing ensures that participated entities cannot deny their actions in the claim submission, approval, and acknowledgment phases.  By engaging all relevant entities in the claim processing, the proposed approach aims to significantly prevent and reduce fraud. It also enhances trust among the stakeholders across the industry, promising substantial benefits for insurers and policyholders by lowering instances of fraud and streamlining insurance claim processing operations.

Looking ahead, future work would focus on developing an end-to-end prototype to test the feasibility of real-time, blockchain-based claim processing in the healthcare ecosystem. This includes evaluating the system's scalability to support a wide category of users and conducting a detailed cost-benefit analysis to assess the economic impacts and viability of the implementation. Furthermore, to improve accessibility and user engagement, it is important to integrate the implemented prototype with mobile applications as users are familiar with the devices and apps.  These initiatives will be instrumental in paving the way for a more efficient and transparent claim process that effectively prevents and minimizes fraud. 

\section*{Acknowledgements} \vspace{-.4em}
This work was partially supported by the U.S. National Science Foundation under Grant No. 1822118 and 2226232. Any opinions, findings, and conclusions or recommendations expressed in this material are those of the authors and do not necessarily reflect the views of the National Science Foundation, or other federal agencies.

\bibliographystyle{IEEEtran}
\bibliography{main}
\end{document}